# The formation of boron sheet at the Ag(111) surface: From clusters, ribbons, to monolayers

Shao-Gang Xu[ab], Yu-Jun Zhao[a], Ji-Hai Liao[a], Xiao-Bao Yang[*a], and Hu Xu[*b]

[a]Department of Physics, South China University of Technology, Guangzhou 510640, People's Republic of China
[b] Department of Physics, South University of Science and Technology of China, Shenzhen, P. R. China

**ABSTRACT**

Boron (B) sheet has been intently studied and various candidates with vacancies have been proposed by theoretical investigations, including the possible growth on metal surface. However, a recent experiment (Science 350, 1513, 2015) reported that the sheet formed on the Ag(111) surface was a buckled triangular lattice without vacancy. Our calculations combined with High-Throughput screening and the first-principles method demonstrate a novel growth mechanism of boron sheet from clusters, ribbons, to monolayers, where the B-Ag interaction is dominant in the nucleation of boron nanostructures. We have found that the simulated STM image of the sheet with 1/6 vacancies in a stripe pattern is in better agreement with the experimental observation, which is energetically favored during the nucleation and growth.

[*]E-mail address: scxbyang@scut.edu.cn
[*]E-mail address: xu.h@sustc.edu.cn

Due to the multi-center bonds, boron nanostructures have shown a striking evolution as thesize increases, which have attracted both theoretical and experimental attentions in the past decades[1-4]. Several planar or quasi-planar structures($B_{17\sim24}$) with tetragonal or pentagonal defects have been determined in the experiments[1],while$B_{30}$and $B_{36}$with the hexagonal vacancy can be considered as the precursor of boron sheet[4]. Stable boron sheets[5, 6]have been found to be a triangular lattice with proper vacancies, leading to extensive theoretical searches which focus on the concentration and distribution of vacancies[7-11].The 2D boron structure with non-zero thickness has been predicted to possess a distorted Dirac cone, with great potential of applications similar to graphene and silicene[12].

The multi-center bonds highly depend on the atomic coordination, which might induce the dramatic effect on the structural stabilities of boron nanostructures. Theoretical investigations have predicted that possible boron sheets could be synthesized on the metal surface (e.g. Cu, Ag, and Au), indicating that the stable sheet with specific concentration and distribution of vacancies would change with the substrate [13-15]. Experimentally, thin boron films were found on the Cu surface with a mixture of boron and boron oxide powders as the boron source, where the monolayer of $B_{28}$ comprises of $B_{12}$icosahedrons and $B_2$dumbbells[16].Notably, the boron sheet with one atom-thick was experimentally determined on the Ag(111) surface using a solid boron atomic source. This simple boron sheet was suggested to be a buckled triangular lattice with no vacancy[17], which is in contrast to the previous theoretically predicted stable configurations of boron sheet on the Ag(111) surface [15].

To give a better understanding of the boron sheet's growth, in this paper we have theoretically investigated the possible B nucleation on the Ag(111) surface based on the first-principle calculations. Surprisingly, we found that small B clusters ($B_{1~3}$) deposited on the silver surface would penetrate into the second layer without energy barrier and further expel Ag atoms on top to form Ag vacancies, similar to the case of silicene growth on the Ag(111) surface[18]. Our simulation demonstrates a clear picture of boron sheet's growth from clusters, ribbons to monolayers, where the B-Ag interaction is dominant in the nucleation of boron clusters. The growth of triangular sheet with 1/6 vacancies in a stripe pattern is found to be most stable and the corresponding STM image is in better agreement with the experimental observation.

The first-principles calculations of B nucleation and growth were based on the density functional theory (DFT) implemented in the Vienna *ab* initio simulation package (VASP)method[19, 20]. The climbing image nudged elastic band method (CI-NEB) is used to search the energy barriers and the reaction pathways[21].Based on the high-throughput screening with a congruence check[22], various configurations of boron nanostructures are considered to model the nucleation on the Ag(111) surface, where the formation energies $E_{form}$ are defined to describe the structural stabilities of boron nanostructures.All these calculation details are shown in Supplementary Material.

To simulate a deposit of the solid atomic boron source on the Ag(111) surface, we considered the adsorption of boron clusters with increasing sizes. Firstly, a single B atom is found to penetrate the first layer with no barrier with the $E_{form}$ of 2.02 eV (c.f. Fig. S1) .The B atom is initially located at 2 Å above the Ag(111) surface, and the total energy gradually decreases as the height decrease. The adsorption and penetration is shown with the energy profile with the distance between the B atom and the substrate, where the CI-NEB calculations also confirmed that the barrier of B atomic penetration approach zero (c.f. Fig. S2). Similarly, the $B_2$ dimer adsorption would also prefer to penetrate through and be located at the second layer, with the $E_{form}$ of 1.51 eV. As shown in Fig.1(a), the $B_2$ dimer would first draw the surface Ag atom up, and lean on the surface. After the penetration, the Ag atom near the $B_2$ dimer is expelled out of its initial position, with the height difference of over 1 Å. When there is one Ag vacancy, the $E_{form}$ of $B_2$ is reduced to 1.45 eV.

Even the triangular $B_3$ determined in experiments would prefer to penetrate through the first layer with the $E_{form}$ of 1.44 eV, where three Ag atoms nearby are upturned with the height difference of around 0.5 Å. $E_{form}$ of $B_3$ is reduced from 1.44 eV to 1.04 eV, when there is one Ag vacancy. Thus, we can expect that the initial nucleation should happen under the first layer, since the bonding between Ag and B atoms is important to stabilize the clusters. Note that Ag atoms would be expelled as more B atoms adsorb on the surface. As shown in Fig.1(b), the barrier of one Ag atom expelled out to the surface is only 0.12 eV when there is a $B_3$ triangle is embedded in the first layer. This Ag atom would further diffuse away on the surface easily, since the diffusion barrier is only 0.10 eV (shown in the inset of Fig.1(b)).

As the number of B atoms increases, the nucleation and growth would be complicated due to the diversity of boron clusters, where there are still controversy on the stable sheet and ribbon of boron on the Ag(111) surface. In practical, we have considered possible triangular fragments to model the small B clusters (shown in Fig. S3), which have been confirmed to be good candidates for the planar stable boron nanostructures. We found that the most stable $B_4$ is of rhombus shape with the $E_{form}$ of 1.25 eV for the adsorption on Ag(111) surface, while it is only 1.03 eV for inside the first layer with one Ag vacancy. The $E_{form}$ of $B_5$ adsorption is 1.09 eV on the Ag surface, while it is 1.02 eV and 0.96 eV for the Ag surface with one and two vacancies respectively. The $E_{form}$ of $B_6$ embedded in the first layer is only 0.66 eV, while it is 0.99 eV for the adsorption on Ag(111) surface. Thus, we show that the Ag atoms would be gradually expelled and the small B clusters would be nucleated inside the first layer of Ag atoms.

Compared to the $B_{7-9}$ clusters (shown in Fig. S4), $B_6$ is a magic structure, where the boron atoms maintain the triangular lattice with the B-Ag bond length of 2.3 A. According to $B_4$ and $B_5$, it is found that the chain with boron triangles would be formed as more Ag atoms are expelled by B atoms. For the Ag(111) surface with a line of vacancies, we found a small ribbon of $B_3$ would be stable with the $E_{form}$ of 0.67 eV. We can also extend $B_6$ to form a ribbon and found the $E_{form}$ is 0.63 eV for the Ag(111) surface with two lines of vacancies. In such structures, boron form the triangular lattice and the B-Ag bond lengths maintain around 2.3 A.

Due to the small mismatch of lattice, boron triangular ribbons can be formed inside the Ag surface with small distortion. We constructed a ribbon of $B_9$ on the Ag surface with two lines of vacancies (shown in Fig. S5) and found that the structure was significantly buckled with a $E_{form}$ of 0.58 eV. For comparison, we constructed a ribbon of $B_9$ with 1/9 vacancies in a strip pattern, which was found to be stable with a $E_{form}$ of only 0.55 eV. For comparison, we also constructed the ribbon from the alpha-sheet and the $E_{form}$ is found to be 0.57 eV. Thus, the ribbon with vacancies in a stripe pattern is energetically preferable during the growth. We have similarly extended the width of this kind of ribbon with the number of line vacancies on the Ag surface increases, where the $E_{form}$ gradually decreases for these two kinds of ribbons. Note that the ribbons with vacancies in a stripe pattern are more stable than that from the alpha-sheet with the same width.

During the nucleation, we found that the edge of Ag surface is dominant to the structural stabilities, where most the B-Ag bond lengths maintain around 2.3 Ås. The triangular lattice of boron ribbon would form gradually with less lattice mismatch. As the width of ribbon increases, vacancies should be necessary to enhance the stability. The vacancies in a stripe pattern are energetically preferable, due to the constraint of symmetry. Thus, the distribution of vacancies with other patterns (such as the alpha sheet) might be not stable. Figure 3 shows the variation of $E_{form}$ as a function of the width, which confirms that the ribbons with vacancies in a stripe pattern should be the most probable in the growth.

In addition, we have considered the ribbon adsorption on the Ag surface without vacancies for comparison. As shown in Fig. 3, the ribbons inside the surface are more stable for small width, and the energy difference decreases gradually as the width increases, since the energy contribution of edge decreases with the size increment. Note that the Ag-B interaction at the edge is strong enough to upturn Ag atoms (shown in Fig. S6), where the configurations are similar to that of embedded structures. We can expect that the growth should along the edge of Ag surface, where more Ag atoms are expelled for the expansion of boron ribbons.

As verified above, the boron ribbons with vacancies in a stripe pattern should be more stable and probable in the growth. For the case of sheet, we found that it is 0.1 eV more stable than that of triangular lattice. Figure 4 shows the simulated STM images of these two sheets for compassion. The parameters used are the same with that in the Ref. 17 and the STM image of buckled triangular lattice is in agreement with that from the Ref. 17. Note that the STM image for the sheet with 1/6 vacancies in a stripe pattern is in better agreement with the experimental results. Recent studies also confirmed the boron sheet with stripe vacancies based on experimental and theoretical methods[23].

In summary, we have theoretically investigated the formation of boron sheet on the Ag(111) surface. Our calculations show that small B clusters ($B_{1\sim3}$) deposited on the metal surface would penetrate the first layer and further expel Ag atoms for extra B atoms. Our simulation demonstrates a clear picture of boron sheet's growth from clusters, ribbons, to monolayers, where the vacancies in a stripe pattern are stable due to the strong B-Ag interaction. We conclude that the boron sheet on the Ag(111) surface should contain 1/6 vacancies in a stripe pattern, which is more energetically preferable and the corresponding STM image is in better agreement with the experimental observation.

**Acknowledgements**

This work was supported by NSFC (Grant Nos. 11474100), Guangdong Natural Science Funds for Distinguished Young Scholars (Grant No. 2014A030306024), and the Fundamental Research Funds for the Central Universities (2015PT017).

**Reference**

[1] A. P. Sergeeva, I. A. Popov, Z. A. Piazza, W. L. Li, C. Romanescu, L. S. Wang, and A. I. Boldyrev, Acc. Chem. Res. **47**, 1349 (2014).
[2] E. Oger, N. R. Crawford, R. Kelting, P. Weis, M. M. Kappes, and R. Ahlrichs, Angew. Chem. Int. Ed. **46**, 8503 (2007).
[3] N. Gonzalez Szwacki, A. Sadrzadeh, and B. I. Yakobson, Phys. Rev. Lett. **98**, 166804 (2007).
[4] Z. A. Piazza, H. S. Hu, W. L. Li, Y. F. Zhao, J. Li, and L. S. Wang, Nat. Commun. **5**, 3113 (2014).
[5] H. Tang and S. Ismail-Beigi, Phys. Rev. Lett. **99**, 115501 (2007).
[6] X. Yang, Y. Ding, and J. Ni, Phys. Rev. B **77**, 041402 (2008).
[7] E. S. Penev, S. Bhowmick, A. Sadrzadeh, and B. I. Yakobson, Nano Lett. **12**, 2441 (2012).
[8] X. Yu, L. Li, X.-W. Xu, and C.-C. Tang, J. Phys. Chem. C **116**, 20075 (2012).
[9] X. Wu, J. Dai, Y. Zhao, Z. Zhuo, J. Yang, and X. C. Zeng, ACS Nano **6**, 7443 (2012).
[10] H. Tang and S. Ismail-Beigi, Phys. Rev. B **82**, 115412 (2010).
[11] H. Lu, Y. Mu, H. Bai, Q. Chen, and S.-D. Li, J. Chem. Phys. **138**, 024701 (2013).
[12] X.-F. Zhou, X. Dong, A. R. Oganov, Q. Zhu, Y. Tian, and H.-T. Wang, Phys. Rev. Lett. **112**, 085502 (2014).
[13] H. Liu, J. Gao, and J. Zhao, Sci. Rep. **3**, 3238 (2013).
[14] Y. Liu, E. S. Penev, and B. I. Yakobson, Angew. Chem. Int. Ed. **52**, 3156 (2013).
[15] Z. Zhang, Y. Yang, G. Gao, and B. I. Yakobson, Angew. Chem. Int. Ed. **53**, 13022 (2015).
[16] G. Tai, T. Hu, Y. Zhou, X. Wang, J. Kong, T. Zeng, Y. You, and Q. Wang, Angew. Chem. Int. Ed. **127**, 15693 (2015).
[17] A. J. Mannix *et al.*, Science **350**, 1513 (2015).
[18] M. Satta, S. Colonna, R. Flammini, A. Cricenti, and F. Ronci, Phys. Rev. Lett. **115**, 026102 (2015).
[19] G. Kresse and J. Hafner, Phys. Rev. B **47**, 558 (1993).
[20] G. Kresse and J. Furthmüller, Phys. Rev. B **54**, 11169 (1996).
[21] G. Henkelman, B. P. Uberuaga, and H. Jónsson, J. Chem. Phys. **113**, 9901 (2000).
[22] S.-G. Xu, Y.-J. Zhao, J.-H. Liao, and X.-B. Yang, J. Chem. Phys. **142**, 214307 (2015).
[23] B. Feng *et al.*, arXiv:1512.05270 [cond-mat.mtrl-sci] (2015).

**Figure Captions**

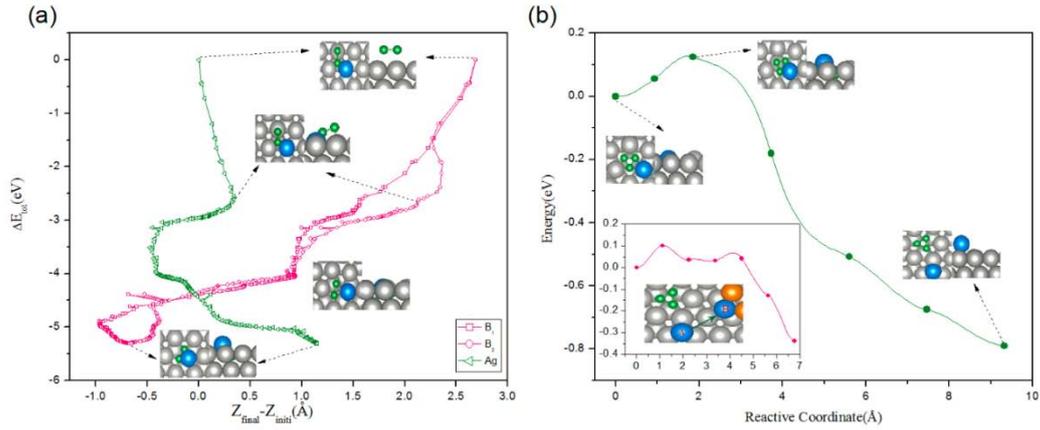

FIG. 1: (Color online) Small boron clusters on the Ag(111) surface. (a) Energy profiles of $B_2$ penetration as a function of the Ag/B heights, where green/grey atoms are for B/Ag atoms and the expelled Ag atom is marked in blue;(b) The barrier of expelling out one Ag atom to the surface and inset is for the Ag atom diffusion on the surface with $B_3$ embedded in the first layer.

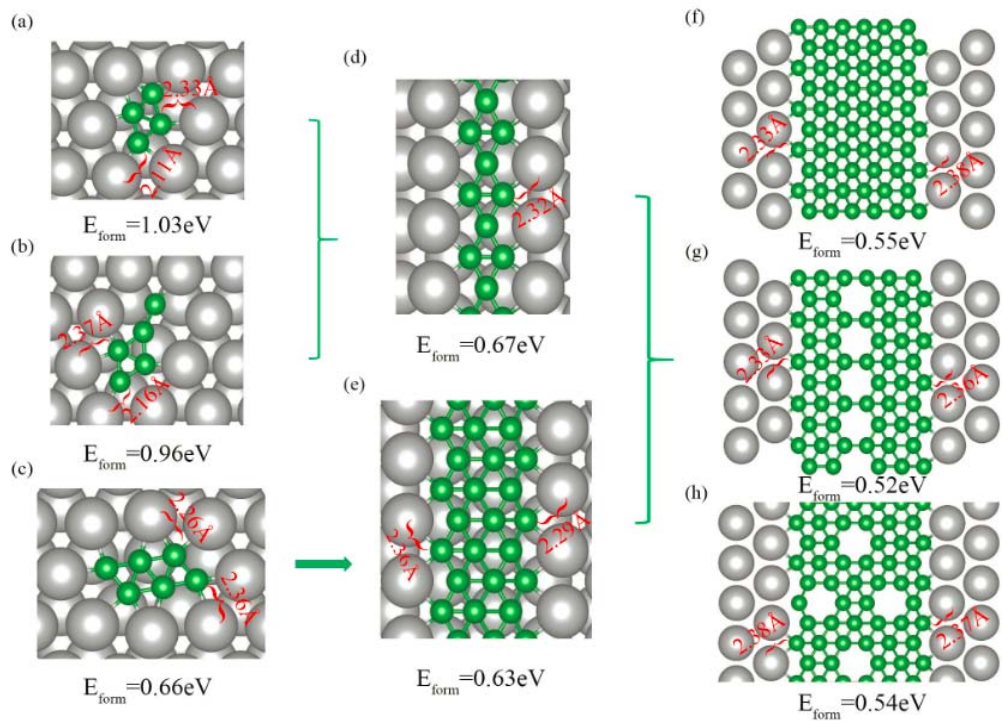

FIG. 2: (Color online) Small boron clusters and the possible ribbons embedded in the Ag(111) surface, where green/grey atoms are for B/Ag atoms.

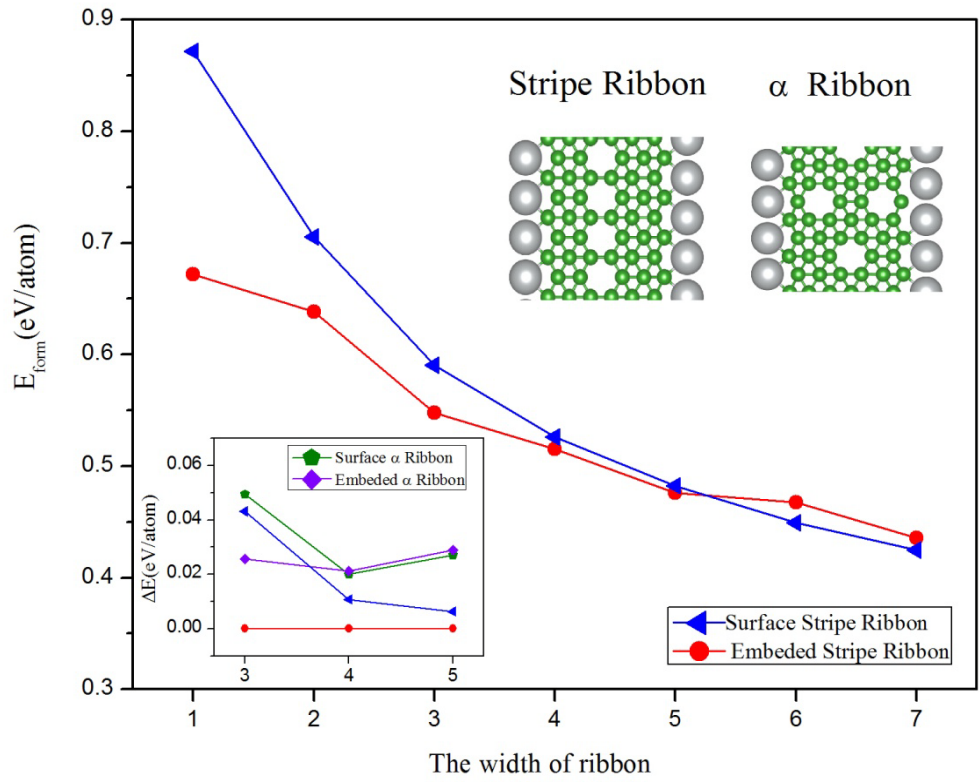

FIG. 3: (Color online) The growth of boron sheet and the energies as a function of width.

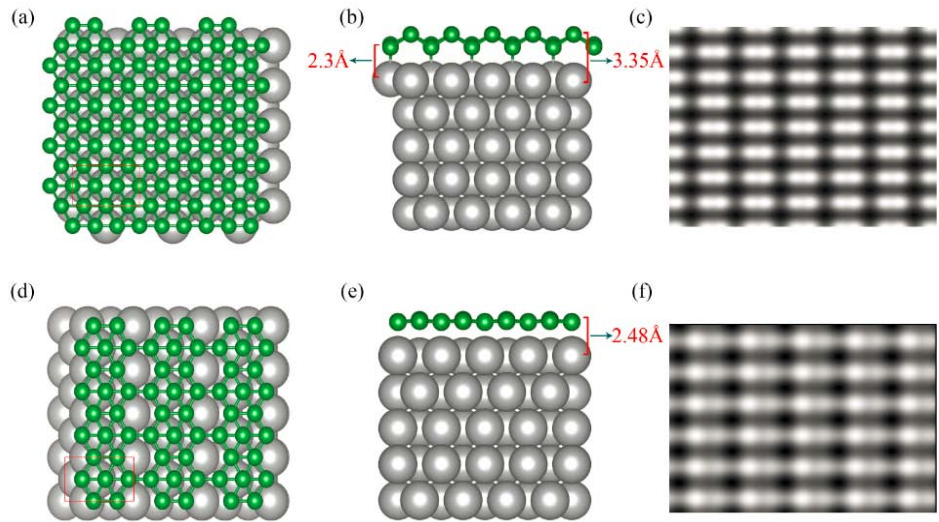

FIG. 4: (Color online) Boron sheets and the corresponding simulated empty states STM images (Vsample = 1.0 V),with overlaid atomic structure.

# Supplemental Materials

Shao-Gang Xu[ab], Yu-Jun Zhao[a], Ji-Hai Liao[a], Xiao-Bao Yang[*a], and Hu Xu[*b]

[a] Department of Physics, South China University of Technology, Guangzhou 510640, People's Republic of China

[b] Department of Physics, South University of Science and Technology of China, Shenzhen, P. R. China

The first-principles calculations of B nucleation and growth were based on the density functional theory (DFT) implementedin the Vienna *ab* initio simulation package (VASP)method[1, 2]. The electron-ion interactions were described by the projector augmented wave (PAW)potentials[3]. To treat the exchange-correlation interaction of electrons, we chose the Perdew-Burke-Ernzerhof (PBE) functional within the generalized-gradient approximation (GGA)[4].All structures are fully relaxed until the force on each atom is smaller than-0.02 eV/Å with the cutoff energy of 450eV. For B clusters on substrates, 6×6 unit cell of Ag(111) surface is used to model the B cluster nucleation, and 3×3×1 Monkhorst-Pack k-points are used. With fixed supercell parameters, the five-layer slab model was further relaxed with thebottom two layers fixed. To avoid interactions between adjacent periodicimages, a vacuum space of more than 14 Å was included in the slab model.The climbing image nudgedelastic band method (CI-NEB) is used to search the energy barriersand the reaction pathways[5]. All these parameters are fully tested to make sure adequate convergence of the calculations.

To describe the stability of a boron monolayer or cluster on Ag(111) surface, we define its formation energy per boron atom as:

$$E_{form} = \frac{1}{n}(E_{tot} - E_{sub} - n \times E_B) \qquad (1)$$

where$E_{sub}$ is the energy of substrate, $E_{tot}$ is the total energy of boron monolayer on Ag(111) system, $E_B$ is the energy per atom inthe boron solid of α phase,*n* is the number of boron atoms in boron monolayer or$B_n$ cluster.

[*] E-mail address: scxbyang@scut.edu.cn
[*] E-mail address: xu.h@sustc.edu.cn

While, if the B nano-ribbon or $B_n$ cluster embedded in the Ag(111) first layer, we define its formation energy per boron atom as:

$$E_{form} = \frac{1}{n}(E_{tot} + m \times E_{Ag} - E_{sub} - n \times E_B) \quad (2)$$

The $E_{Ag}$ is energy per atom of the edge atoms on the Ag(111), $m$ is number of the expelled Ag atoms, all these atoms are diffused to the edge of the Ag(111), all the other variables are the same as the definitions in Eq. (1).

Fig. S1 Stable structures of $B_{1-3}$ adsorption and embedded in the Ag(111) surface.

$E_{form}$=2.02 eV

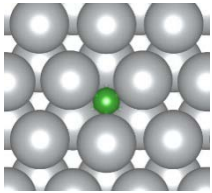

$E_{form}$=1.51 eV

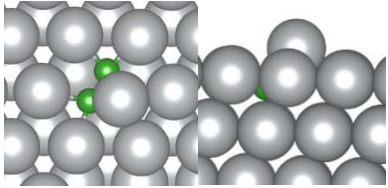

$E_{form}$=1.45 eV

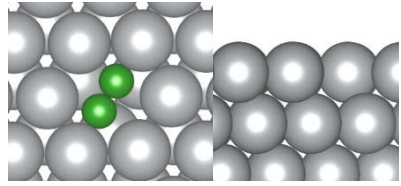

E_form=1.44 eV

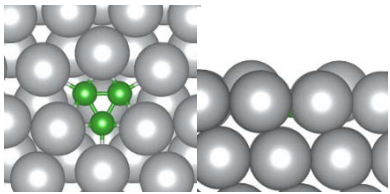

E_form=1.04 eV

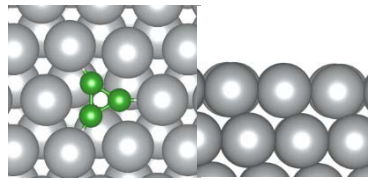

Fig. S2 The energy profile and barrier for $B_{1-3}$ penetrating the first layer of Ag(111) surface.

B1(Energy Profile)                                    NEB

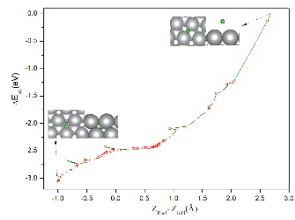
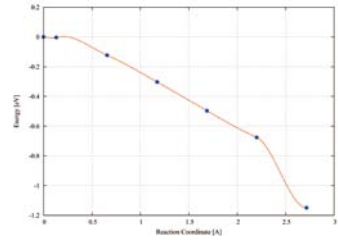

B2(Energy Profile)　　　　　　　　　　　　　　NEB

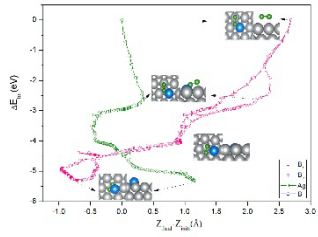
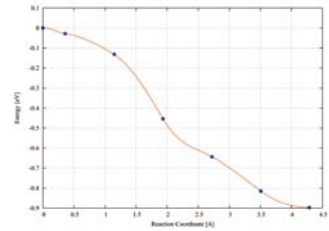

B3(Energy Profile)　　　　　　　　　　　　　　NEB

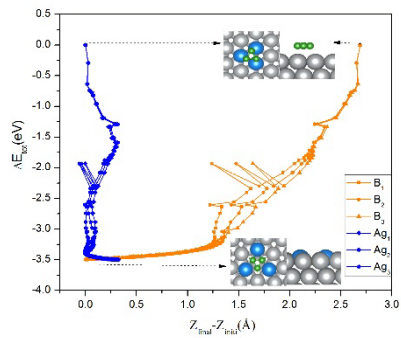
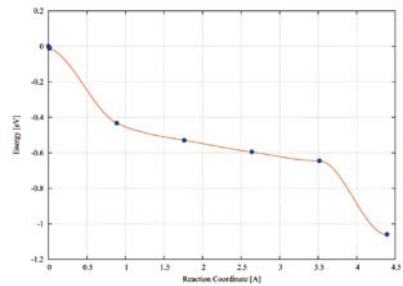

As the number of B atoms increases, there are a lot of possible structures. According to our previous study, the triangular fragments can be modeled as the small B clusters. The clusters of $B_{4\sim9}$ are found to be more stable with more nearest bonds, while the structures could be hardly stable if there are atoms with only one nearest neighbor. Based on the high-throughput screening with a congruencecheck, we have calculated the interaction of about 200 B clusters with the metal substrate, including various configurations of the adsorption on metal surface and that embedded in the first layer. After the full relaxation, the stable structures with various sizes become similar, though the initial configurations might be quite different.

Fig. S3 The triangular candidates for B$_{5-8}$ adsorption on the Ag(111) surface.

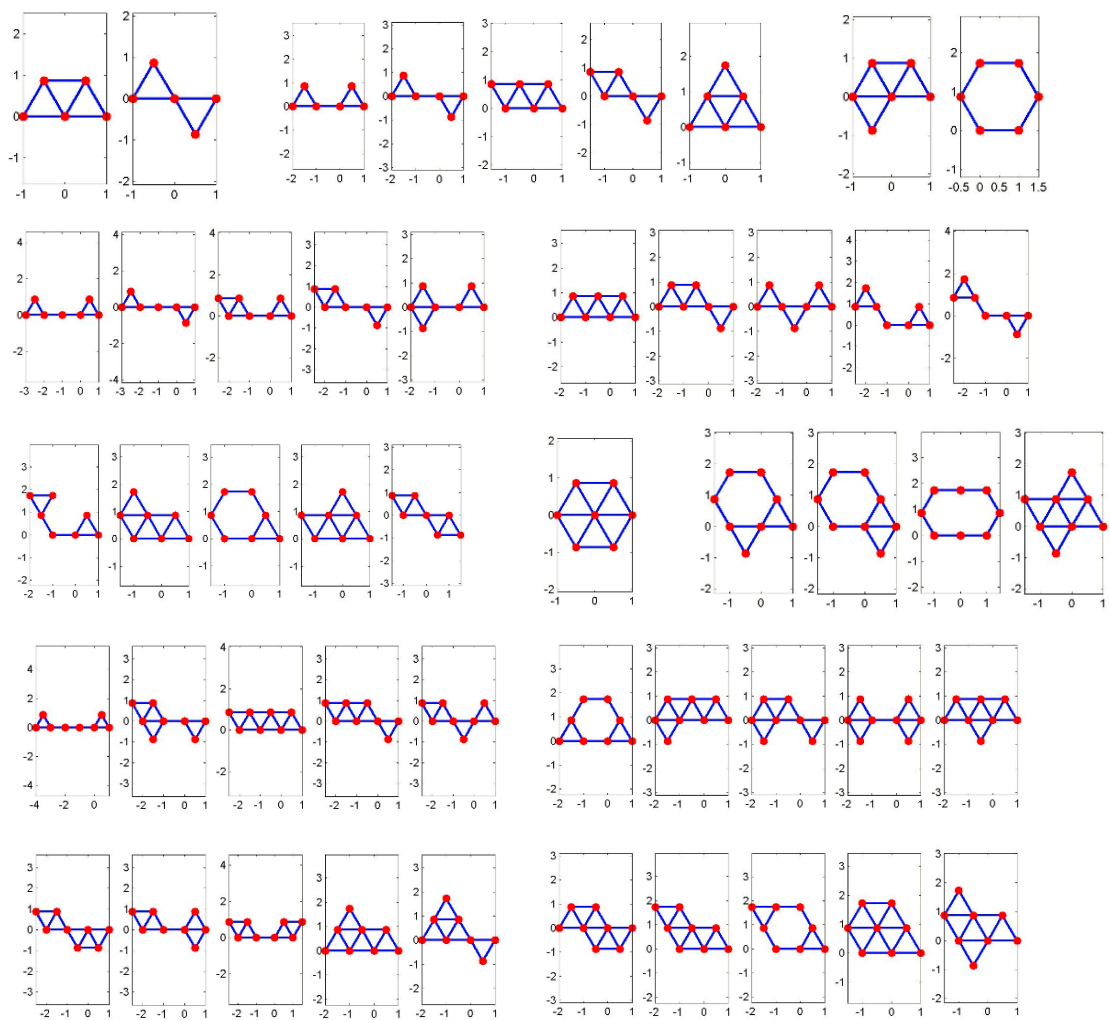

Similarly, the B$_{7-9}$ clusters are more stable to be embedded in the first layer, instead of adsorption. The configurations and formation energies are shown as follows:

Fig. S4 Stable B$_{7-9}$ embedded in the Ag(111) surface.

E$_{form}$=0.82 eV                                  E$_{form}$=0.78 eV

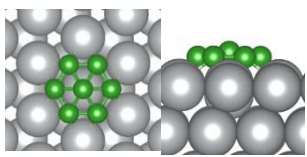                                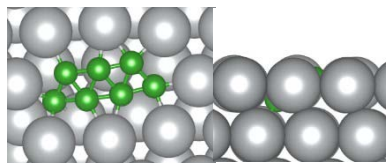

E$_{form}$=0.83eV                                   E$_{form}$=0.80 eV

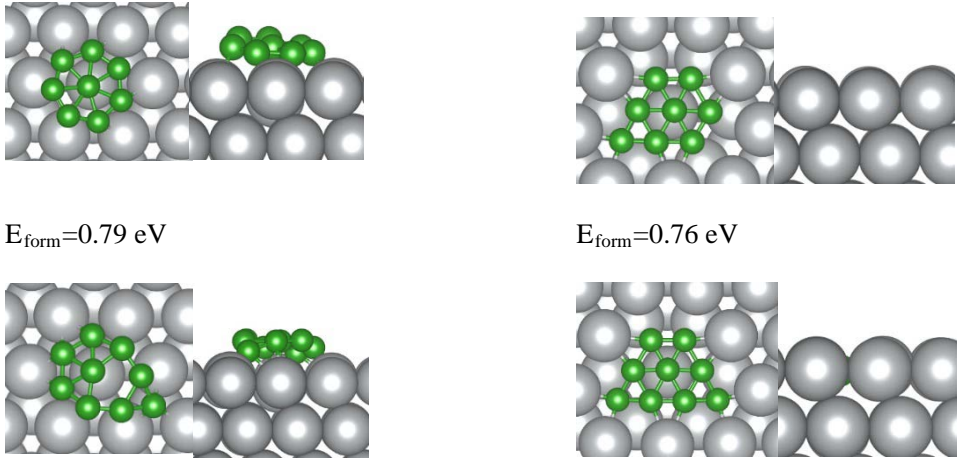

$E_{form}$=0.79 eV

$E_{form}$=0.76 eV

Fig. S5. Possible ribbons structures during the growth.

V(Ag3)

$E_{form}$=0.58eV   $E_{form}$=0.55eV   $E_{form}$=0.57 eV

V(Ag4)

$E_{form}$=0.55eV   $E_{form}$=0.52eV   $E_{form}$=0.54eV

V(Ag5)

$E_{form}$=0.53eV   $E_{form}$=0.48eV   $E_{form}$=0.51 eV

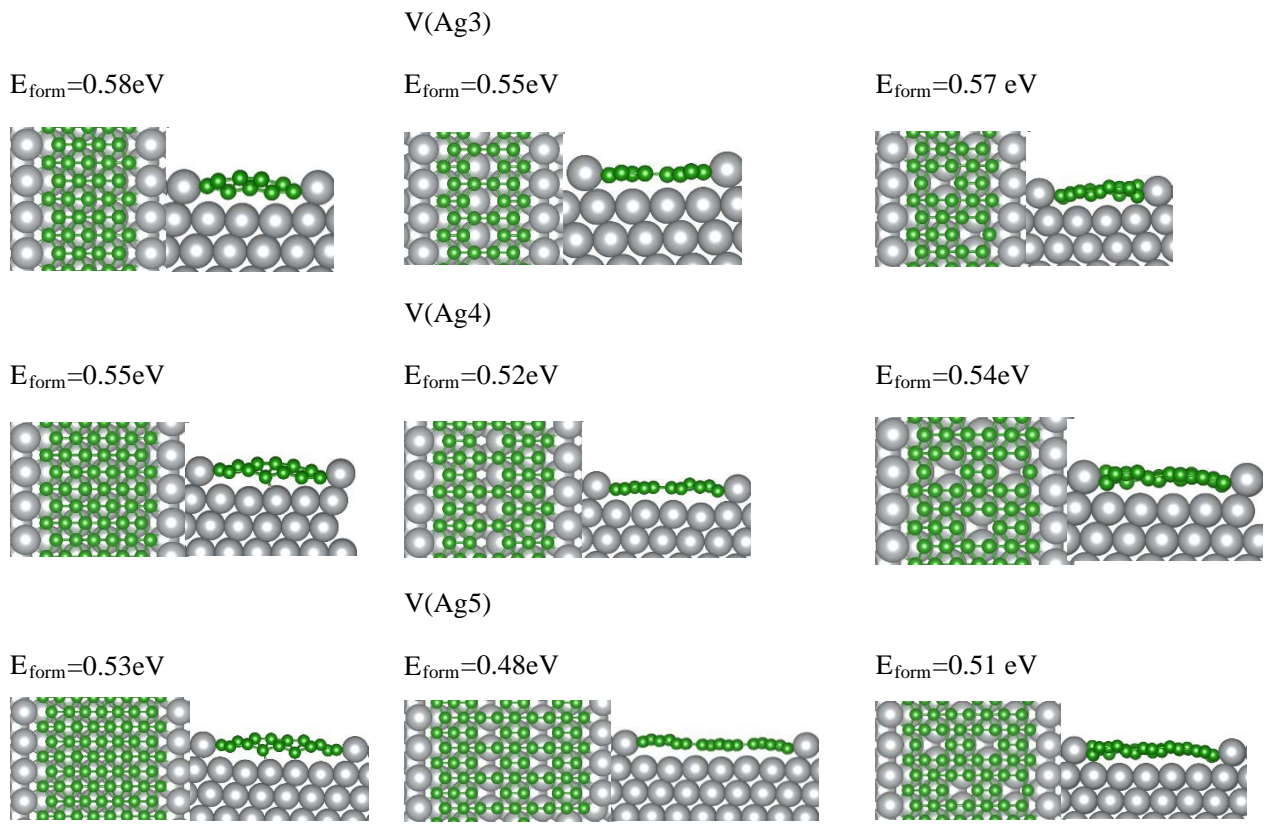

Fig. S6. Embedded structures and ribbon adsorption on surface.

Adsorption on surface                Embedded structures

V(Ag1)

$E_{form}$=0.87eV
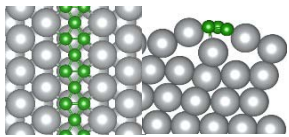

$E_{form}$=0.67eV
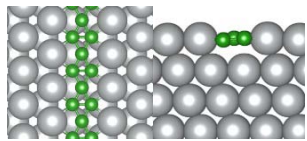

V(Ag2)

$E_{form}$=0.71eV
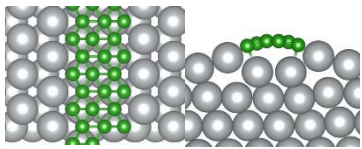

$E_{form}$=0.63eV
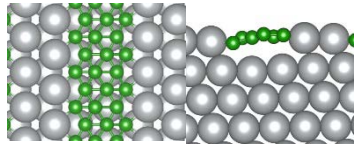

V(Ag3)

$E_{form}$=0.59eV
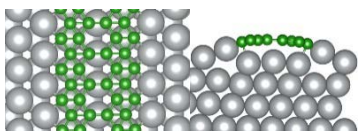

$E_{form}$=0.55eV
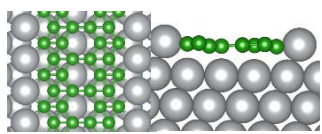

V(Ag4)

$E_{form}$=0.53eV
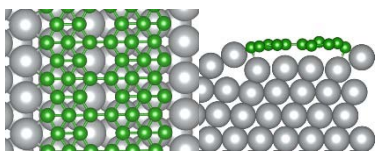

$E_{form}$=0.51eV
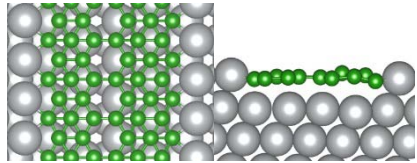

V(Ag5)

$E_{form}$=0.48eV
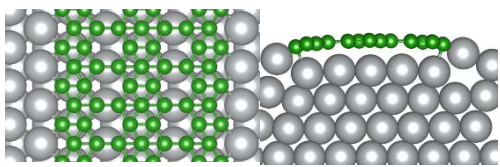

$E_{form}$=0.47eV
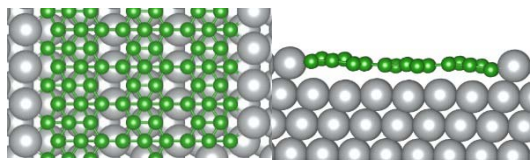

V(Ag6)

$E_{form}$=0.45eV
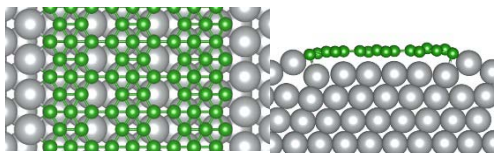

$E_{form}$=0.46eV
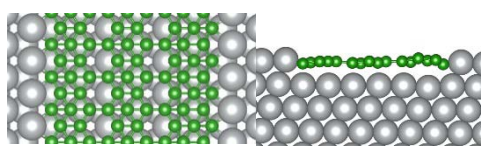

V(Ag7)

$E_{form}$=0.42eV
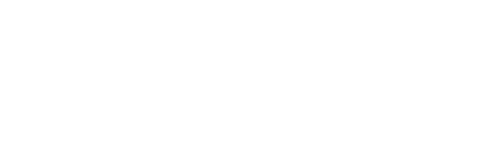

$E_{form}$=0.43eV
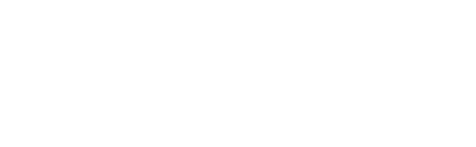

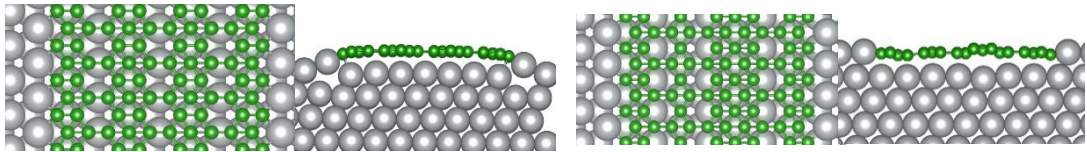

Reference

[1]  G. Kresse and J. Hafner, Phys. Rev. B **47**, 558 (1993).
[2]  G. Kresse and J. Furthmüller, Phys. Rev. B **54**, 11169 (1996).
[3]  G. Kresse and D. Joubert, Phys. Rev. B **59**, 1758 (1999).
[4]  J. P. Perdew, K. Burke, and M. Ernzerhof, Phys. Rev. Lett. **78**, 1396 (1997).
[5]  G. Henkelman, B. P. Uberuaga, and H. Jónsson, J. Chem. Phys. **113**, 9901 (2000).